\documentclass[prd,superscriptaddress,nofootinbib,amsmath,amssymb,aps,11pt]{revtex4-2}
\usepackage[utf8]{inputenc}

\usepackage{bigints}
\usepackage{bm}
\usepackage{amsfonts}
\usepackage{latexsym}
\usepackage{graphicx}
\usepackage{amsmath}
\usepackage{anyfontsize}
\usepackage{palatino}
\usepackage{mathpazo}
\usepackage[normalem]{ulem}
\usepackage{soul}
\usepackage{xcolor}

\usepackage{xcolor}     

\begin{document}
\title{ Effects of mass and self-interaction on nonlinear scalarization of scalar-Gauss-Bonnet black holes}

\author{Alexandre M. Pombo}
\email{pombo@fzu.cz}
\affiliation{CEICO, Institute of Physics of the Czech Academy of Sciences, Na Slovance 2, 182 21 Praha 8, Czechia}

\author{Daniela D. Doneva}
\email{daniela.doneva@uni-tuebingen.de}
\affiliation{Theoretical Astrophysics, Eberhard Karls University of T\"ubingen, T\"ubingen 72076, Germany}
\affiliation{INRNE - Bulgarian Academy of Sciences, 1784 Sofia, Bulgaria}

\begin{abstract}
    It was recently found that in certain flavours of scalar-Gauss-Bonnet gravity linearly stable bald black holes can co-exist with stable scalarized solutions. The transition between both can be ignited by a large nonlinear perturbation, thus the process was dubbed nonlinear scalarization, and it happens with a jump that leads to interesting astrophysical implications. Generalizing these results to the case of nonzero scalar field potential is important because a massive self-interacting scalar field can have interesting theoretical and observational consequences, e.g. reconcile scalar-Gauss-Bonnet gravity with binary pulsar observation, stabilize black hole solutions, etc. That is why in the present paper, we address this open problem. We pay special attention to the influence of a scalar field mass and self-interaction on the existence of scalarized phases and the presence of a jump between stable bald and hairy back holes. Our results show that both the addition of a mass and positive self-interaction of the scalar field result in suppression or quenching of the overall scalarization phenomena. A negative scalar field self-interaction results in an increase of the scalarization. The presence and the size of the jump, though, are not so sensitive to the scalar field potential. 
\end{abstract}

\maketitle
%

%
\section{Introduction}
%
    According to the Kerr hypothesis, the astrophysical black holes that we observe in the Universe are described by the famous Kerr metric in general relativity (GR). Thus, they are characterized solely by their mass and spin. A strong argument in support of this conjecture is not only the validity of Einstein's theory of gravity in various observations but also the proof of a number of uniqueness theorems for electro-vacuum \cite{ruffini1971introducing,chrusciel2012stationary,robinson1975uniqueness,carter1971axisymmetric,bekenstein1995novel,israel1968event} (see also \cite{chrusciel2012stationary} for a review). According to them, a Kerr black hole is also a solution in a number of modified theories of gravity. There are different ways to go outside of the validity of these theorems \cite{Herdeiro:2015waa} (see e.g. \cite{herdeiro2018spontaneous,herdeiro2019black,cardoso2013black,cardoso2013matter,ramazanouglu2018regularization,herdeiro2014kerr,Stefanov:2007eq,Doneva:2010ke}). One of the interesting and well-motivated ones is to consider black holes in an effective field theory involving higher curvature invariants such as the quantum gravity motivated scalar-Gauss-Bonnet (sGB) theory \cite{kanti1996dilatonic,Torii:1996yi,Pani:2009wy,Sotiriou:2013qea}. Because of the presence of a scalar field coupled to the Gauss-Bonnet invariant, hairy black hole can exist in this case.

    As it turns out, there are different ways to ignite the scalar field in sGB gravity depending on the properties of the scalar field coupling function. The most common case is the shift-symmetric sGB or Einstein-dilaton-Gauss-Bonnet theory where a scalar field is always present around the black holes \cite{kanti1996dilatonic,Torii:1996yi,kanti1998dilatonic,gross1987quartic,Kleihaus:2011tg}. A second interesting option is the case of spontaneous scalarization \cite{doneva2018new,silva2018spontaneous,antoniou2018evasion} when the GR black hole is always a solution of the sGB field equations but it becomes linearly unstable below a certain black hole mass giving rise to a stable scalarized solution\footnote{ Similar studies were performed for electrically charged BHs~\cite{brihaye2019spontaneous,jiang2020spontaneous}, spinning BHs \cite{hod2020onset,dima2020spin,herdeiro2021spin,berti2021spin,collodel2020spinning}, spinning and charged BH~\cite{herdeiro2021aspects,annulli2022spin}; and with a vector field instead of a scalar field \cite{barton2021spontaneously}.}. In that case, a number of observational constraints can be elegantly circumvented in sGB gravity because it practically coincides with GR in the weak field regime. Below we will call this case ``normal scalarization''. It is interesting that another type of scalarization can also exist when the Schwarzschild black hole is always a linearly stable solution within sGB theory but stable scalarized black holes can exist as well \cite{doneva2022beyond} (for a similar phenomenon in Einstein-Maxwell-scalar gravity see \cite{blazquez2020einstein,blazquez2021quasinormal}). Approximate rotating nonlinearly scalarized black holes were also constructed \cite{Lai:2023gwe,Doneva:2022yqu}.  Such scalarized black hole phases are thermodynamically preferred over Schwarzschild for a large range of the parameter space and the scalar field around them can be ignited only through a large nonlinear perturbation of a Schwarzschild black hole. Thus, we will call this phenomenon ``nonlinear scalarization''. A mixture between the normal scalarization and the nonlinear one can also exist. In that case, Schwarzschild black hole is unstable below a certain mass but still, in a certain region of the parameter space, both the bald and the hairy linearly stable solutions can exist. In the presence of nonlinearly scalarized phases a jump between the two stable (non-scalarized and scalarized) black hole branches can happen that has very intriguing astrophysical implications \cite{Doneva:2022byd}.  Interestingly, a similar phenomenon can be observed for neutron stars as well \cite{Doneva:2023kkz}.

    The above-mentioned studies in sGB gravity consider the simpler case of a zero scalar field potential. Non-vanishing scalar field mass or self-interaction can also have very interesting effects. For example, it can suppress the scalar dipole emission acting as an effective screening mechanism \cite{Ramazanoglu:2016kul} and reconciling the theory e.g. with the binary pulsar observations \cite{Danchev:2021tew}. On a theory level, a self-interaction term can stabilize otherwise unstable black hole solutions \cite{Minamitsuji:2018xde,Macedo:2019sem,Silva:2018qhn}. Compact objects in sGB gravity with nonzero scalar field mass were considered also in \cite{Hod:2019vut,Doneva:2019vuh,Bakopoulos:2020dfg,Xu:2021kfh,Peng:2020znl}. Nonzero potential in the context of nonlinear scalarization was not considered until now. Such a study is particularly interesting since in some cases the nonlinear scalarized phases are detached from the bald Schwarzschild solution. It is important to investigate the existence of hairy black holes in that case and to check whether the presence of a jump between the different phases, with the related astrophysical manifestations,  still survives for a strong enough scalar field mass or self-interaction. This is exactly the focus of the present paper.

    Throughout the paper, $4\pi G=1=4\pi\epsilon_0$. The signature of the spacetime is $(-,+,+,+)$. In this work, one is solely interested in spherical symmetry and the metric matter functions are only radially dependent. For notation simplicity, after being first introduced, the functions' radial dependence is omitted, e.g. $X(r)\equiv X$, and  $X' \equiv dX/dr$, and we consider the notation $X_{,\phi}\equiv dX /d\phi$ for the derivative with respect to the scalar field.

    The paper is organized as follows. In Sec.~\ref{Sec.Frame} we introduce the model's action as well as the metric ansatz and self-interaction potential that allows us to obtain the field equations in Sec.~\ref{Sec.Fieldeq}. The coupling function between the scalar field and the Gauss-Bonnet term is introduced in Sec.~\ref{Sec.Coup}. The proper boundary conditions are imposed in Sec.~\ref{Sec.Fieldeq}, letting us obtain a set of illustrative results of nonlinear scalarization and simultaneous linear and nonlinear scalarization, Sec.~\ref{Sec.Num}. Results are shown for a massive scalar field, Sec.~\ref{Sec.mass}, and in the presence of a quartic self-interaction Sec.~\ref{Sec.Inter}. We end the manuscript with the conclusion of our results Sec.~\ref{Sec.Conc}. 
%
\section{Framework}\label{Sec.Frame}
%
	The action in scalar-Gauss-Bonnet gravity with a scalar field minimally coupled to the Gauss-Bonnet invariant and a non-vanishing potential $U(\phi)$, is defined by the action
    \begin{equation}
        S=\frac{1}{16\pi}\int d^4x \, \sqrt{-g} \Big[R-2\nabla _\mu \phi \nabla ^\mu \phi + \lambda ^2 f(\phi)R_{GB}^2 -U(\phi)\Big]\ ,
    \end{equation}
    where $R$ is the Ricci scalar with respect to the spacetime metric $g_{\mu \nu}$ and the real scalar field, $\phi$, is non-minimally coupled to the Gauss-Bonnet invariant, $R_{GB}^2$, through a dimensionless coupling function $f(\phi)$; $\lambda$ is the so-called Gauss-Bonnet coupling constant that has dimensions of length. The Gauss-Bonnet invariant comes as
    \begin{equation}
        R_{GB} ^2 = R^2 -4R_{\mu \nu}R^{\mu \nu}+R_{\mu \nu \alpha \beta}R^{\mu \nu \alpha \beta}\ .
    \end{equation}
	The system's field equations are given by the Einstein-Klein-Gordon system with the Gauss-Bonnet term
    \begin{eqnarray}\label{eq:FE}
        &R_{\mu \nu}-\frac{1}{2}g_{\mu \nu}R+\Gamma _{\mu \nu}=2\nabla_\mu \phi \nabla _\nu \phi -g_{\mu \nu}\nabla_\alpha \phi \nabla ^\alpha \phi -\frac{g_{\mu \nu}}{2}U(\phi)\ , \label{eq:FE1} \\
        &\nabla_\alpha \nabla ^\alpha \phi = \frac{1}{4}\frac{dU(\phi)}{d\phi}-\frac{\lambda ^2}{4}\frac{df(\phi)}{d\phi}R_{GB}^2\ \label{eq:FE2} .
    \end{eqnarray}
    The tensor $\Gamma_{\mu \nu}$ that modifies the left-hand side of the Einstein equations is defined as
    \begin{align}
        \Gamma _{\mu \nu} = &R \big( \nabla_\mu \psi _\nu +\nabla_\nu \psi _\mu\big)-4 \nabla^\alpha \psi_\alpha \Big(R_{\mu \nu}-\frac{1}{2}R\, g_{\mu \nu}\Big)+4R_{\mu \nu} \nabla ^\alpha \psi_\nu \nonumber\\
       & +4 R_{\nu \alpha}\nabla^\alpha \psi_\mu-4g_{\mu \nu}R^{\alpha \beta}\nabla_\alpha \psi_\beta + 4 R^\beta _{\ \mu \alpha \nu}\nabla^\alpha \psi_\beta \ ,
    \end{align}
    with
    \begin{equation}
        \psi_\mu = \lambda ^2 \frac{df(\phi)}{d\phi}\nabla_\mu \phi\ .
    \end{equation}

    For the line element, let us consider a standard metric ansatz that is compatible with a static spherically symmetric spacetime and contains two unknown functions
    \begin{equation}\label{eq:metric}
        ds^2 = -\sigma ^2 (r) N(r) dt^2+\frac{dr^2}{N(r)}+r^2\big(d\theta^2+\sin ^2 \theta d\varphi ^2\big)\ ,\ \ {\rm with}\ \ N(r)=1-\frac{2m(r)}{r}\ ,
    \end{equation}
    where $m(r)$ is the Misner-Sharp mass function~\cite{misner1964relativistic} and $\sigma(r)$ is an unknown metric function\footnote{Note that $\delta=\log(\sigma\sqrt{N})$ is the redshift function.}. The scalar field possesses the same symmetry as the spacetime and hence is solely radially dependent, i.e. $\phi (t,r,\theta,\varphi)\equiv \phi (r)$. The latter is under a self-interaction potential containing a mass and a quartic self-interaction term\footnote{As one will see ahead (Sec. \ref{Sec.Num}), the scalar field amplitude is always smaller than unity (having the maximum at the horizon), resulting in a decreased impact of higher-order polynomial terms to the potential.}
    \begin{equation}\label{eq:potential}
        U(\phi)=\mu ^2 \phi ^2+\beta \phi ^4 \ .
    \end{equation}
%
    \subsection{Coupling function}\label{Sec.Coup}
%
    In this work, we would like to study the impact of a mass or a quartic self-interaction term on the nonlinear scalarization studied in \cite{doneva2022beyond}. For that purpose, one should design the coupling function $f(\phi)$ properly. The first requirement is that the GR black holes should also be solutions within sGB gravity. After examining the field eqs. \eqref{eq:FE1} and \eqref{eq:FE2}, one can easily conclude that this can be secured by the requirement 
    \begin{equation}
        \frac{df}{d\phi}\Big|_{\phi = 0} = 0\ .
    \end{equation}
    The second derivative of $f(\phi)$ on the other hand controls the type of scalarization, i.e. whether it is ``normal'' scalarization, nonlinear one, or a mixture of both. In the first and the third case, we should have $\frac{d^2f}{d\phi ^2}\Big|_{\phi = 0} > 0$ (for a Schwarzschild solution), making the GR black holes linearly stable only if they are massive enough. The pure nonlinear scalarization occurs in the absence of tachyonic instabilities when the Schwarzschild solution is always linearly stable against linear scalar perturbations, i.e. for $\frac{d^2f}{d\phi ^2}\Big|_{\phi = 0} = 0$.

    The condition for nonlinear scalarization is easily satisfied if the leading order term in the expansion of $f(\phi)$ is at least cubic in  $\phi$. In this work, we focus on $Z_2$ symmetry theories and thus we will employ the following function:
	\begin{equation}
	 f_1 (\phi) =\frac{1-e^{k\phi ^4}}{4k}\ .
	\end{equation}	    
    We have chosen an exponential form of the coupling function instead of a polynomial because it often leads to better numerical behaviour -- scalarized solutions exist for a larger range of the parameter space -- and at least one of the branches is linearly stable.
    
    In addition, we are also interested in models that contain simultaneously linear and nonlinear instability, which we will define as \textit{mixed} models. For this reason, let us consider the additional exponential coupling:
    \begin{equation}
	 f_2 (\phi) = \frac{1- e^{-b(\phi^2 + k\phi^4)}}{2b}\ .
    \end{equation}
    As one can see, it satisfies the condition for ``normal'' scalarization  $\frac{d^2f}{d\phi ^2}\Big|_{\phi = 0} > 0$ that leads to destabilization of small-mass Schwarzschild black holes. The parameter $b$ is set to $b=6$, however, further values of $b$ are possible and known to originate similar results (see \cite{doneva2018charged} for e deeper discussion). The quartic term in $\phi$, though, allows for the co-existence of linearly stable bald GR and scalarized phases in a certain region of the parameter space, similar to pure nonlinear scalarization.
%
	\subsection{Field equations}\label{Sec.Fieldeq}
%
   Assuming static spherically symmetric spacetime and scalar field configuration, the field equations reduce to two first order (for $m$ and $\sigma$) and one second order (for $\phi$) coupled ordinary differential equations
	\begin{align}\label{Eom1}
      m'  = & \frac{1}{4 \left(r^3-4 \lambda ^2 r (r-3 m) \phi ' f_{,\phi}\right) } \Big[ 16 \lambda ^2 r\, m (r-2 m) \phi '' f_{,\phi}+2 \phi ' \Big(8 \lambda ^2 r\, m (r-2 m) \phi ' f_{,\phi} \nonumber\\
        &-8 \lambda ^2 r\, m f_{,\phi}+24 \lambda ^2 m^2 f_{,\phi}-2 r^4 m \phi '+r^5 \phi'\Big)-r^5 U\Big]\ , \\
       \sigma'  = & \frac{\sigma}{r^2-4 \lambda ^2 (r-3 m) \phi ' f_{,\phi}}\Big[\phi'^2 \left(4 \lambda ^2 m f_{,\phi}+r^3\right)+4 \lambda ^2 m \phi '' f_{,\phi}\Big]\ ,\\
        \phi ''  = & -\frac{1}{4 r^5 (r-2 m) \sigma }\Bigg\{r^2 \Bigg[4 r \sigma ' \left(r^3 \phi '-4 \lambda ^2 m' f_{,\phi}\right)+\sigma \Big(16 \lambda ^2 m'^2 f_{,\phi}-8 r^3 \left(m'-1\right) \phi'\Big)\Bigg] \nonumber\\
        &+r^4 U_{,\phi}-8 r m \Bigg[2 \lambda ^2 r f_{,\phi} \Big(r \sigma ''-\left(5 m'+1\right) \sigma '\Big)+\sigma \Big(2 \lambda ^2 f_{,\phi}\left(4 m'-r m''\right)+r^3 \phi '\Big)+r^4 \sigma ' \phi '\Bigg]\nonumber \\
        &+16 \lambda ^2 m^2 f_{,\phi} \Big(r \left(2 r \sigma ''-5 \sigma '\right)+3 \sigma \Big)\Bigg\}\ .\label{Eom3}
    \end{align}
	To solve the set of $3$ ODEs, one must impose proper boundary conditions. At infinity, asymptotical flatness is guaranteed by imposing 
    \begin{equation}\label{Inf1}
        \phi \sim \frac{Q_s}{r}e^{-r\mu}\ , \qquad \qquad \sigma \sim 1-\frac{Q_s ^2}{2r^2}\ , \qquad \qquad m\sim M-\frac{Q_s ^2}{2r}\Big(1+2\lambda^2 U_{,\phi}f_{,\phi}\Big)\ ,
    \end{equation}
    where $Q_s$ is the scalar charge. 
    While at the horizon, the functions can be approximated by a polynomial series expansion in $(r-r_H)$
    \begin{align}\label{Hor1}
    &m\approx \frac{r_H}{2}+\Bigg[\frac{1}{2}-\frac{r_H \big(2+r_H ^2 U\big)}{4\big(r_H+2\phi_1\lambda ^2f_{,\phi}\big)}\Bigg](r-r_H)+ \mathcal{O}\left((r-r_H)^2\right)\ ,\nonumber\\
    &\sigma \approx \sigma _0 +\sigma_0\, \phi_1 ^2\frac{r_H^2 +2\lambda^2 f_{,\phi}}{r_H+2\phi_1 \lambda^2f_{,\phi}}(r-r_H)+ \mathcal{O}\left((r-r_H)^2\right)\ ,\\
    &\phi \approx \phi_0+\phi _1 (r-r_H)+ \mathcal{O}\left((r-r_H)^2\right)\ ,\nonumber
    \end{align}
    where
    \begin{align}
    &\phi_1= -\frac{1}{4 \lambda ^2 f_{,\phi} \Big[U \big(r_H^4-4 \lambda ^4 f_{,\phi}^2\big)+r_H^2 \big(\lambda ^2 f_{,\phi} U_{,\phi}+2\big)\Big]}\Bigg[\big(r_H ^5-12 \lambda ^4 r_H f_{,\phi}^2\big)U +2 \lambda ^2 r_H ^3 f_{,\phi} U_{,\phi}-2 \lambda ^4 r_H^3 U^2 f_{,\phi}^2\nonumber\\
    &+\sqrt{(r_H ^2 U+2)^2 +\Big(-12 \lambda ^6 r_H ^2 f_{,\phi}^3 U_{,\phi}+4 \lambda ^8 U f_{,\phi}^4 (r_H^2 U+12)-8 \lambda ^4 r_H^2 f_{,\phi}^2 \left(r_H^2 U+3\right)+r_H^6\Big)}+2 r_H^3\Bigg]\ . \nonumber
    \end{align}
    The last equation for $\phi_1$ involves a square root and thus regular black hole solutions exist only when the term under the root is positive.
%
    \subsection{Identities and physical quantities of interest}\label{Sec.Ident}
%
    When the instability settles, an additional class of solutions besides the vacuum ones emerges. These are the scalarized solutions that we are interested in. Besides the ADM mass, $M$, these solutions are also characterized by the so-called ``scalar charge'' $Q_s$, which, however, is not associated with a conservation law but comes from the radial decay behaviour of the scalar field. There are also a number of relevant horizon quantities: the Hawking temperature $T_H$, the horizon area $A_H$, and the entropy $S_H$. The black hole's entropy is the sum of two terms
    \begin{equation}
        S_H = S_{EH}+S_{sGB}\ , \quad {\rm with}\quad S_{EH} = \frac{A_H}{4}\ , \quad S_{sGB} = \frac{\lambda ^2}{2}\int _H d^2 x \sqrt{h} f(\phi) R^{(2)}\ ,
    \end{equation}
    where $R^{(2)}$ is the Ricci scalar of the induced horizon metric $h$. The solutions satisfy a Smarr law
    \begin{equation}\label{Sma}
     M = 2 T_H S_H + M_s\ ,
    \end{equation}
    where $M_s$ is the contribution of the scalar field 
    \begin{equation}
      M_s = \frac{1}{2} \int d^3 g \sqrt{-g} \big(\partial _a \phi \big) ^2\ .
    \end{equation}
    Also, the solutions satisfy the first law of black hole thermodynamics
    \begin{equation}
      dM = T_H dS_H\ ,
    \end{equation}
    in which there is no contribution from the scalar field.

	At last, the solutions obey the so-called virial identity \cite{herdeiro2022deconstructing,herdeiro2021virial,hunter2022novel,derrick1964comments}
    \begin{align}\label{Vir}
        \bigintss _{r_H} ^{\infty} dr \ \Bigg\{& r\big( 3r-2r_H\big)U\, \sigma+2\phi' \Bigg[\frac{8}{r^4} \lambda^2 m f_{,\phi}\bigg(\sigma\Big[ 3\big(r-r_H\big)m+r\big(3r-2r_H\big)m'\Big]+\nonumber\\
       & \sigma'r\Big[r\big(-2r+r_H\big)+ m\big(6r-4r_H\big)\Big]\bigg)-\Big(r(r-2r_H)+2r_Hm\Big)\sigma\phi'\Bigg]\Bigg\} = 0\ .
    \end{align}
\bigskip
   Observe that the model's equations  are invariant under the transformation 
    \begin{equation}\label{scaling}
        r \to \alpha r\ , \qquad \qquad \lambda \to \alpha \lambda\ ,
    \end{equation}
    with $r$ the radial coordinate and $\alpha >0$ an arbitrary positive constant. Following standard terminology, let us define the reduced quantities,
    \begin{equation}
        a_H \equiv \frac{A_H}{16\pi M^2}\ , \qquad {\rm and} \qquad t_H \equiv 8\pi T_H M \ ,
    \end{equation}
    which will be considered in what follows. In addition, observe that, since $\lambda$ contains dimensions of length, all the other quantities can be scaled accordingly, namely
    \begin{equation}
     \frac{M}{\lambda}\ ,\qquad \frac{r_H}{\lambda}\ ,\qquad \frac{S_H}{\lambda ^2}\ ,\qquad \mu \lambda\ ,\qquad \beta\lambda ^2\ .
    \end{equation}

%
\section{Numerical results}\label{Sec.Num}
%
   The field equations \eqref{Eom1}-\eqref{Eom3} together with the boundary conditions at infinity \eqref{Inf1} and at the horizon \eqref{Hor1} previously obtained, form a Dirichlet boundary problem. They are solved using an in-house developed, parallelized, adaptative step-size $6(5)^{th}$ order explicit Runge-Kutta integration method with the boundary conditions being imposed through a secant strategy to the initial scalar field amplitude $\phi _0$ and metric function $\sigma _0$. 
   
   In all solutions, it was guaranteed a maximum local error of $10^{-15}$ during integration, while the boundary conditions were imposed with a tolerance of $10^{-8}$. The physical accuracy was computed through the virial identity \eqref{Vir} and the Smarr relation~\eqref{Sma}, both required to contain a relative difference no larger than $10^{-6}$ and $10^{-3}$, respectively.  

    Below we present solutions for various values of the scalar particle's mass (Sec.~\ref{Sec.mass}) and both positive and negative values of the quartic self-interacting term (Sec.~\ref{Sec.Inter}).
%
    \subsection{Massive scalar field}\label{Sec.mass}
%
    Let us start our analysis with the massive scalar field case while the effect of self-interaction will be left for the following subsection. For this, and following the work done in \cite{doneva2022beyond,doneva2020spin,peng2020spontaneous,doneva2019gauss}, we have considered five exemplary branches:
    \begin{align}
        & {\rm \textit{Nonlinear}}: \qquad  f_1 (\phi) =\frac{1-e^{k\phi ^4}}{4k} \qquad {\rm with} \qquad k=\{25,\, 50,\, 1000\}\ , \label{eq:coupling_nonline}\\
        &  \textit{Mixed}:\qquad 	 f_2 (\phi) = \frac{1- e^{-b(\phi^2 + k\phi^4)}}{2b}\qquad {\rm with} \qquad k =\{4,\,32\}\quad {\rm and}\ \quad  b=6  \ . \label{eq:coupling_mixed}
    \end{align}
   These correspond to the three possible kinds of solutions shown in \cite{doneva2022beyond} for the case of fully nonlinear scalarization and two examples of mixed scalarization. The case with standard scalarization ($k=0$) was already extensively studied in the literature \cite{doneva2018new,doneva2022beyond,herdeiro2021aspects,doneva2018charged,jiang2020spontaneous,brihaye2019spontaneous} including the case of a massive scalar field \cite{Silva:2018qhn,Doneva:2019vuh,Staykov:2021dcj,doneva2020spin}, and hence we will solely focus on the new solutions with nonlinear and mixed scalarization.

    \begin{figure}[h!]
         \centering
          \includegraphics[scale=0.65]{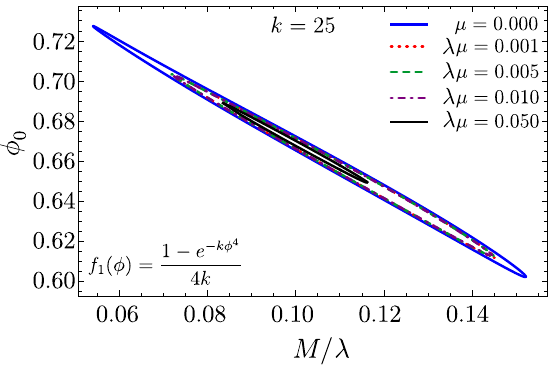}\hfill
          \includegraphics[scale=0.65]{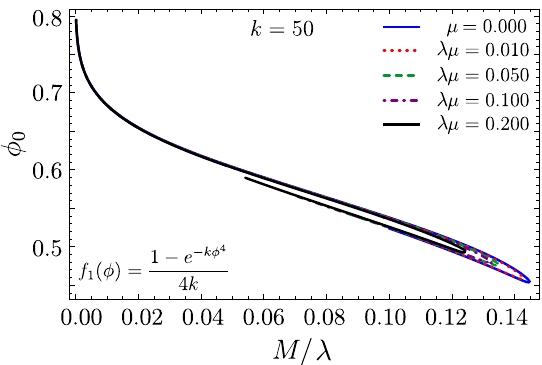}\\
          \includegraphics[scale=0.65]{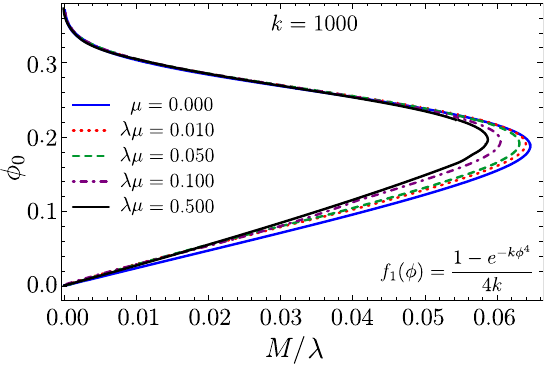}\\
         \includegraphics[scale=0.65]{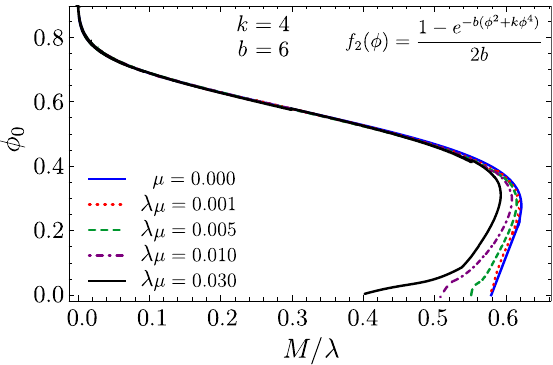}\hfill
        \includegraphics[scale=0.65]{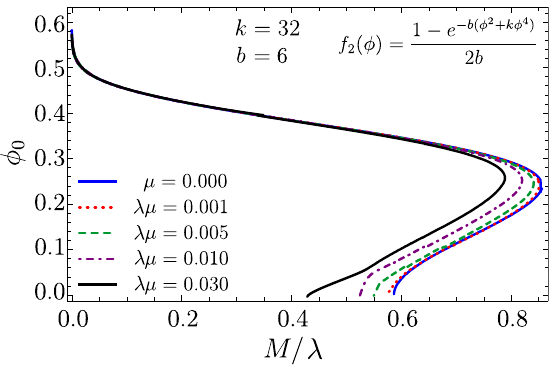}
       \caption{Scalar field amplitude at the horizon, $\phi _0$, as a function of the scaled ADM mass, $M/\lambda$, for several values of the scaled scalar field particle's mass, $\mu \lambda$, and for five different cases of the scalarized GB black holes: nonlinear with coupling function \eqref{eq:coupling_nonline} and $k= 25$ (\textit{top-left}), $k=50$ (\textit{top-right}) and $k=1000$ (\textit{middle}); and mixed coupling function \eqref{eq:coupling_mixed} with $b=6$ and $k=4$ (\textit{bottom-left}) and $k=32$ (\textit{bottom-right}).}
        \label{F1}
    \end{figure} 
  
    Let us start by observing the scalar field amplitude at the horizon, $\phi_0$ (see Fig.~\ref{F1}). The top and the middle rows represent the three distinct structures of solution branches in the case of pure nonlinear scalarization. Namely, for large values of $\kappa$ (e.g. $\kappa=1000$ in the middle row) two branches of solutions exist -- an upper potentially stable one that merges at some maximum mass with a lower unstable branch. For intermediate $\kappa$ (e.g. $\kappa=50$ in the top-right panel), the unstable lower branch starts deforming and it never reaches the $M=0$ limit. For even smaller $\kappa$ like in the top-left panel, the two branches of solutions form a closed loop. The Schwarzschild solution is always linearly stable in these cases. A general observation is that the upper part of the scalarized solution branches (having larger $\phi_0$ for a fixed mass), are also potentially stable, while the lower scalarized branches are always unstable \cite{Blazquez-Salcedo:2022omw}.
    
    The mixed scalarization in the bottom panels of Fig. \ref{F1} is simpler. The branches of solutions start at a bifurcation point of the Schwarzschild $\phi_0=0$ solutions (the point of origin of the scalarized branches on the $x$-axes) followed by an increase of their mass and scalar field until a maximum mass is reached. After that, the mass starts decreasing towards the $M=0$ limit. The part of the branch between the bifurcation point and the maximum mass is always unstable while the rest of the sequence is formed by stable black holes \cite{Blazquez-Salcedo:2022omw}. The Schwarzschild solution on the other hand is stable only for masses larger than the bifurcation point. For a more detailed discussion on the structure of solution branches as well as their stability, we refer the reader to \cite{doneva2022beyond,Blazquez-Salcedo:2022omw} (a similar behaviour also occurs in Einstein-Maxwell-scalar models \cite{blazquez2020einstein,blazquez2021quasinormal}).

    As one can conclude from the figures and the discussion above, in all of the considered cases there is a jump between the stable scalarized black hole branches and the Schwarzschild one that can have interesting astrophysical implications \cite{Doneva:2022byd}. One of our tasks will be to investigate the effect of scalar field potential on the presence and size of this jump.

\begin{figure}[h!]
		 \centering
		 \includegraphics[scale=0.60]{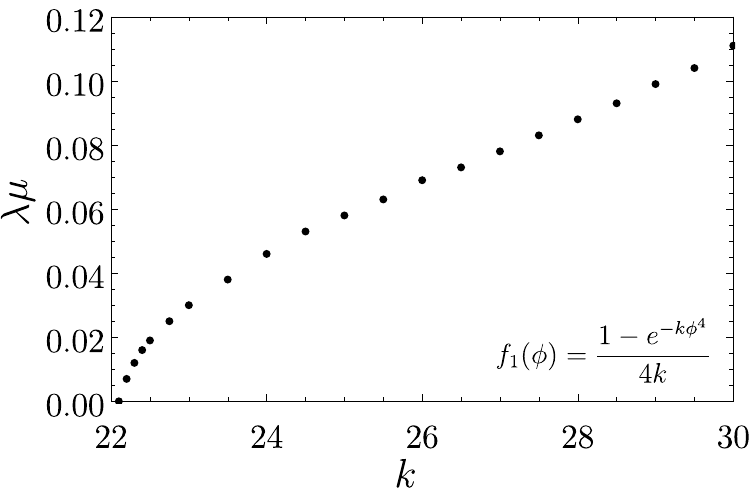}\hfill
          \includegraphics[scale=0.60]{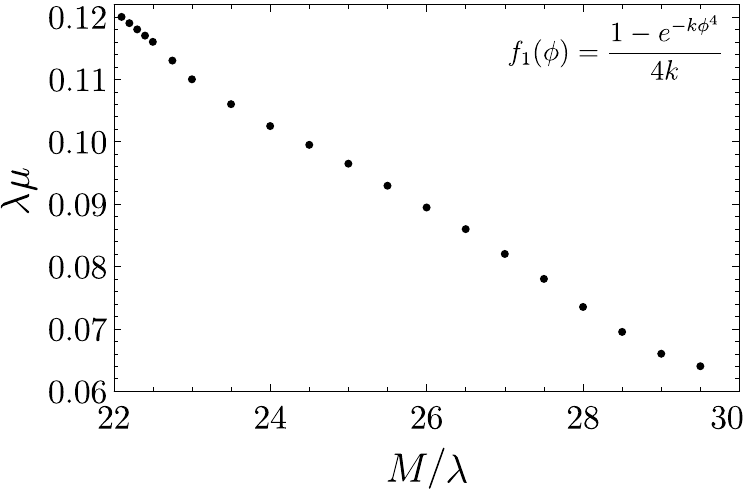}
   \caption{The maximum scaled scalar field particle's mass, $\mu \lambda$, for which nonlinearly scalarized sGB black holes (with coupling function \eqref{eq:coupling_nonline}) still exist, as a function of (\textit{left panel}) the coupling constant, $k$, and (\textit{right panel}) scaled ADM mass, $M/\lambda$. Alternatively, one can interpret the results as the minimum value of the nonlinearly scalarized $k$ (or $M/\lambda$) able to support nonlinear scalarization for a given scalar field particle's mass, $\mu \lambda$. Each plot point corresponds to the configuration at which the close loop of scalarized black hole branches tends to a point (see e.g. $k=25$ in Fig.~\ref{F1} top-left).}
	 	\label{F5}
		\end{figure}

   After discussing in detail the general structure of branches let us turn now to the effects of scalar field mass and self-interaction. Observes that, independently of the branch or kind of scalarization, the presence of a mass term results in a quench of the scalarization phenomena: larger $\mu \lambda$ leads to a smaller $\phi_0\, vs\, M/\lambda$ parameter range for which a given $k$ model solution's exists (the domain of the existence shrinks). As an example, the $k=25$ nonlinear solution's domain of existence reduces to a point and stops existing for $\mu \lambda\approx 0.058$ (see also Fig.~\ref{F5}). While for the $k=32$ mixed scalarization there is a simultaneous decrease of the domain of existence width and a shift of the normal scalarization bifurcation point to higher values of $\lambda$ for the same ADM mass, $M$. Both cases are a clear signature of the quenching of the scalarization phenomena.

    In the case of mixed scalarization (bottom panels of Fig. \ref{F1}), the scalar field mass also shortens the domain of existence of the unstable scalarized branch (from the bifurcation point to the maximum mass). This effectively shrinks the area where stable scalarized branches co-exist with linearly stable Schwarzschild branch holes. The ``height'' of the jump between the two, though, is affected much less by the nonzero $\mu$.  

\begin{figure}[h!]
		 \centering
          \includegraphics[scale=0.65]{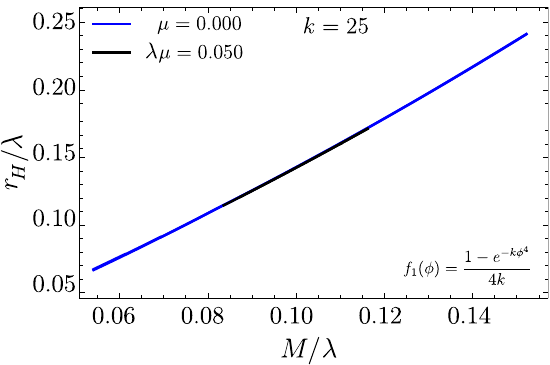}\hfill
          \includegraphics[scale=0.65]{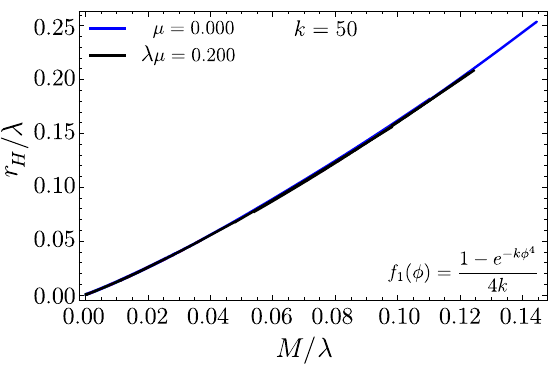}\\
          \includegraphics[scale=0.65]{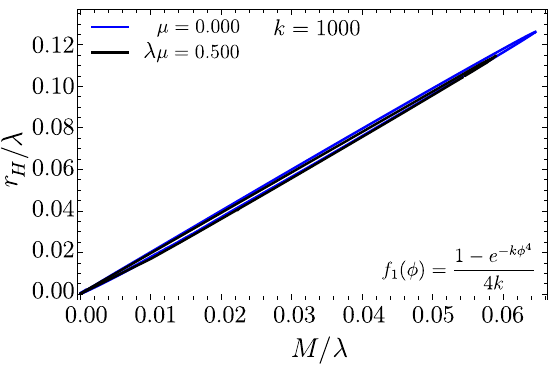}\\
		 \includegraphics[scale=0.65]{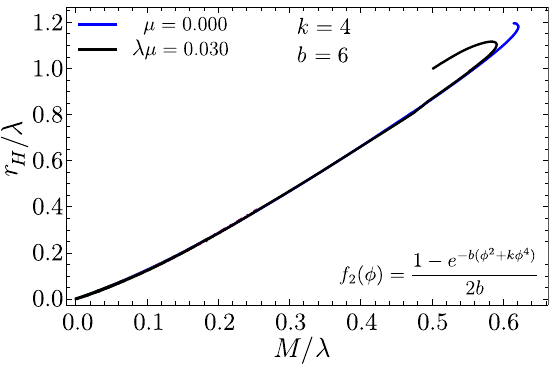}\hfill
   	 \includegraphics[scale=0.65]{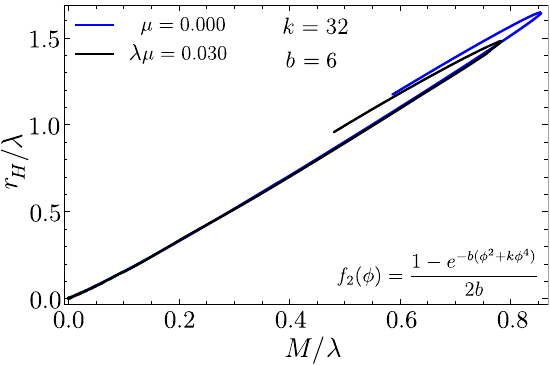}
	   \caption{Black hole's scaled horizon radii, $ r_H/\lambda$, as a function of the scaled ADM mass, $M/\lambda$, for several values of the scaled scalar field particle's mass, $\mu \lambda$, and five different cases of the scalarized GB black holes: nonlinear with coupling function \eqref{eq:coupling_nonline} and $k= 25$ (\textit{top-left}), $k=50$ (\textit{top-right}) and $k=1000$ (\textit{middle}); and mixed coupling function \eqref{eq:coupling_mixed} with $b=6$ and $k=4$ (\textit{bottom-left}) and $k=32$ (\textit{bottom-right}).}
	 	\label{F2}
		\end{figure}

    Concerning the horizon radii, Fig.~\ref{F2}, a similar behaviour can be observed: the possible solution's parameter region shrinks. In particular, there is a decrease of the maximum horizon radii with the increase of the particle's mass. The same behaviour can be seen for the entropy, Appendix~\ref{A}.

    At last, observe the close loop domain of existence of $k=25$, Fig.~\ref{F1} (top-left). The increase of the $\lambda \mu$ term causes the loop to shrink to a point and disappear. This behaviour occurs for all the close loop domains of existence, with the transition happening at somewhat larger $\kappa$ for larger $\lambda \mu$. In particular, the minimum value $k$ for $\mu \lambda =0$ occurs at $k\approx 22.1$; while for $k=25$ the scalarized branches disappear for scalar field masses above $\mu \lambda \approx 0.058$. We have investigated this problem in detail (see  Fig.~\ref{F5}). In the left panel, the points indicate the limiting value of the parameters for which we could find scalarized solutions. Black holes with nontrivial scalar hair exist only below this curve.

    Thus, in the case of nonlinear scalarization, hairy black hole solutions seem to exist for arbitrary large $k$ while the minimum $k$ is $\mu \lambda$ dependent. A detailed investigation of our result hints towards the conclusion that the open branch-like scalarized solutions (see $k=50$ in Fig. \ref{F1}) continuously transform into a close loop-like (see $k=25$ in Fig. \ref{F1}) as one increases $\mu \lambda$ \footnote{A similar transition probably exists when reducing $k$.}. However, the latter was not possible to prove due to numerical difficulties. In addition, Fig.~\ref{F5} right panel, shows the dependency of the limiting solution's $\lambda \mu$ with the BH's mass. A higher particle's mass requires a higher coupling constant $\lambda$ for the same $M$. A massive scalar field is less prone to scalarize and hence requires a higher coupling constant to counter-balance the quenching by the mass term.
%
    \subsection{Massless, self-interacting scalar field}\label{Sec.Inter}
\begin{figure}[h!]
		 \centering
   	 \includegraphics[scale=0.65]{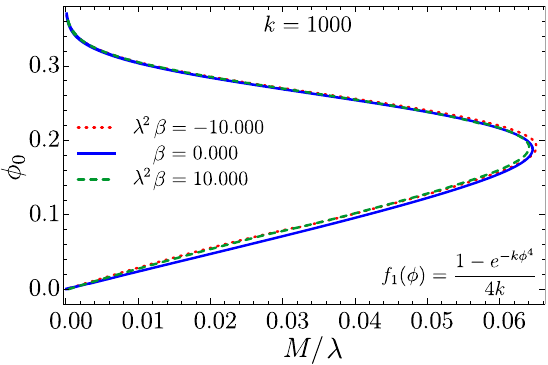}\hfill
          \includegraphics[scale=0.65]{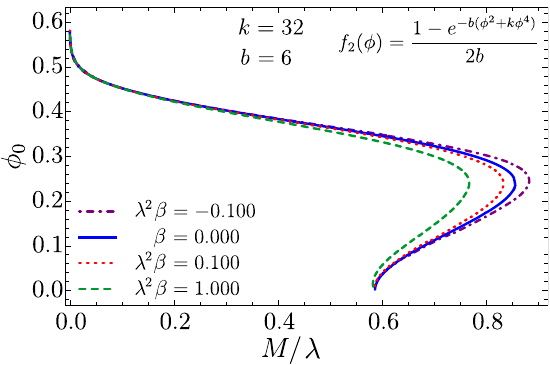}
	   \caption{Scalar field amplitude at the horizon, $\phi _0$, as a function of the scaled ADM mass, $M/\lambda$, for several values of the scaled scalar field quartic self-interaction, $\lambda ^2 \beta$, and two different cases of the scalarized sGB black holes: a nonlinear with coupling function \eqref{eq:coupling_nonline} and $k=1000$ (\textit{left panel}); and a mixed coupling function \eqref{eq:coupling_mixed} with $b=6$ and $k=32$ (\textit{right panel}).}
	 	\label{F6}
		\end{figure}

%
    To solely isolate the impact of the self-interaction, consider the case of a massless, quartic self-interacting scalar field for two examples: the nonlinear scalarization, $f_1(\phi)$, with $k=1000$, and the mixed scalarization, $f_2(\phi)$, with $ k=32$. While in the case of the mass term, only positive values are possible, the self-interaction can be either positive or negative, with a positive/negative $\beta$ in eq. \eqref{eq:potential} decreasing/increasing the domain of existence for which scalarized solutions exist -- see Fig.~\ref{F6} for both nonlinear (left panel) and mixed (right panel) scalarization. We have to note that, from an equation point of view, adding a self-interaction is very similar to adding a quartic term in the coupling, a fact that has already been noticed in the case of standard spontaneous scalarization \cite{Macedo:2019sem}. The reason is that on the right-hand side of the scalar field Klein-Gordon equation \eqref{Eom3}, enter both the coupling function ($f_{,\phi}$) and potential ($U_{,\phi}$) derivates.

   	\begin{figure}[h!]
		 \centering
		 \includegraphics[scale=0.65]{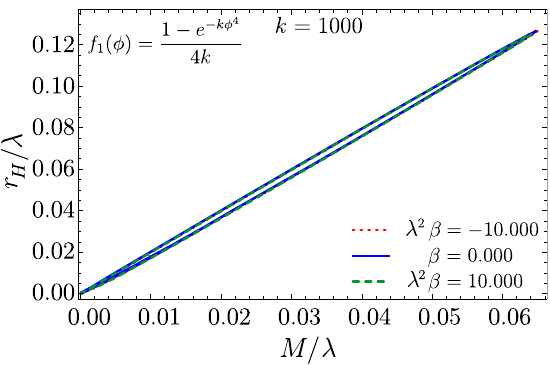}\hfill
          \includegraphics[scale=0.65]{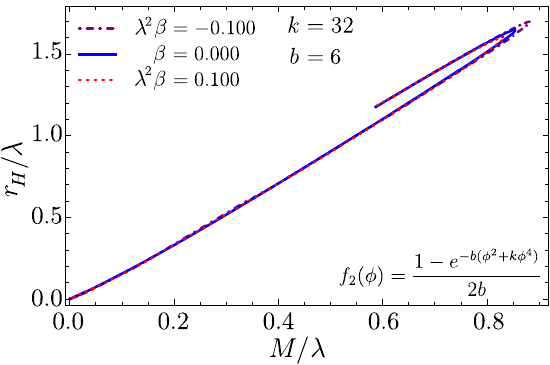}
	   \caption{Black hole's scaled horizon radii, $ r_H/\lambda$, as a function of the scaled ADM mass, $M/\lambda$, for several values of the scaled scalar field particle's quartic self-interaction, $\lambda^ 2 \beta$, and two different cases of the scalarized sGB black holes: a nonlinear with coupling function \eqref{eq:coupling_nonline} and $k=1000$ (\textit{left panel}); and a mixed coupling function \eqref{eq:coupling_mixed} with $b=6$ and $k=32$ (\textit{right panel}).}
	 	\label{F7}
		\end{figure} 
    From the analysis of the domain of existence, Fig.~\ref{F6}, two behaviours are evident. First, the quartic self-interaction has a smaller impact on the overall scalarization than the mass term. This is a natural consequence of the fact that the scalar field maximum value occurs at the horizon, $\phi_0$, and is always less than unity (see Fig.~\ref{F1} and \ref{F6}), i.e. ${\rm max} \big(\phi (r)\big)=\phi _0<1$. Thus, the effect of the quartic (and higher order) self-interaction is weaker when compared with the mass term. 

    Second, the self-interaction, $\lambda^2 \beta$, has less influence on the nonlinear scalarization ($k=1000$, left panel) than on the mixed scalarization ($k=32$, right panel) \footnote{Observe that the same is true also for the mass term, Fig.~\ref{F1}, but up to a smaller extent.}. While for the nonlinear scalarization with $k=1000$, $\lambda ^2\beta =10.0$ barely alters the domain of existence, for the mixed scalarization with $k=32$, even a relatively weak self-interaction like $\lambda ^2\beta = 0.1$ is enough to visible change it. This is a result of the fact that for pure nonlinear scalarization, hairy black holes are typically present for stronger coupling, $\lambda$, between the scalar field and the Gauss-Bonnet invariant (thus smaller $M/\lambda$) compared to the mixed scalarization. In other words, for the same mass, the tachyonic instability in the mixed coupling allows scalarization to occur for weaker couplings (larger $M/\lambda$) than the nonlinear scalarization, making it more sensitive to additional interactions.

	Notice, though, that similar to the massive scalar field term, the presence and the height of the jump between the last stable scalarized solution and the bald GR black holes do not change.

	At last, we have verified numerically that in the case of the mixed scalarization, the bifurcation point from the vacuum solutions does not change with the introduction of the self-interaction. The reason behind this is that the bifurcation point is solely dependent on the linear terms in the $f_{,\phi}$ that enters in the right-hand side of the Klein-Gordon equation \eqref{Eom3}.
%
\section{Conclusion}\label{Sec.Conc}
%
  	In this work, we studied the nonlinear black hole scalarization phenomena due to the presence of a massive or self-interacting scalar field nonminimally coupled to the Gauss-Bonnet invariant. We considered both pure nonlinear scalarization when Schwarzschild is linearly stable but stable hairy black holes can also be present, as well as mixed linear and nonlinear scalarization when Schwarzschild is unstable below a certain mass but still a region of the parameter space exists where linearly stable bald and hairy black holes can co-exist.  
   
    For the pure nonlinear scalarization, we focused on three values of the constant $k$ that define the coupling function between the scalar field and the Gauss-Bonnet term. These three cases cover all the interesting possibilities for the domain of existence of nonlinearly scalarized solutions. In the mixed case, the solution structure is much simpler and two branches with different $k$ were examined. We have observed indications of a suppression of the scalarization phenomena by a mass/positive self-interaction term, which is independent of the scalarization type. In particular, the domain of existence shrinks and moves to higher values of the coupling between the scalar field and the Gauss-Bonnet invariant.
    
    In fact, the mass term is able to cancel the scalarization in the nonlinear case. For a relatively small $k$, the nonlinear scalarization's domain of existence forms a close loop that shrinks with the increase of the scalar field's particle's mass until it becomes a point and vanishes. In the mixed scalarization, on the other hand, there is a shift of the bifurcation point (the point where the Schwarzschild solution becomes unstable giving rise to hairy black hole) to higher values of the coupling constant, however, without ever disappearing.
     
    The magnitude of the effect associated with the mass/self-interaction terms is also different for both scalarization types. The reason is that the tachyonic instability in the mixed coupling makes the black hole more susceptible to scalarization. As a result, the latter requires a weaker coupling of the scalar field to the Gauss-Bonnet invariant to ignite the scalar hair development, when compared to nonlinear scalarization. This makes the mixed scalarized black holes more susceptible to the influence of the self-interaction potential

   What makes the nonlinear and mixed scalarized black holes particularly interesting is the presence of a jump between the last stable scalarized solution and the bald GR. Thus a transition between the two can have interesting astrophysical implications. Our results indicate that even though the existence domain changes sometimes significantly for massive/self-interacting scalar field, the ``height'' of the jump between the two black hole phases is much less sensitive, and at least for the considered value of the parameters it is only very weakly affected.

%
\section*{Acknowledgments}

%
We would like to thank S. Yazadjiev for advice and carefully reading the manuscript. A. M. Pombo is supported by the Czech Grant Agency (GA\^CR) under grant number 21-16583M. D. D. acknowledges financial support via an Emmy Noether Research Group funded by the German Research Foundation (DFG) under Grant No. DO 1771/1-1. The partial support of KP-06-N62/6 from the Bulgarian science fund is also gratefully acknowledged.

\appendix 
\addcontentsline{toc}{section}{APPENDICES}

%
\section{Thermodynamic properties}\label{A}
\begin{figure}[h!]
		 \centering
          \includegraphics[scale=0.65]{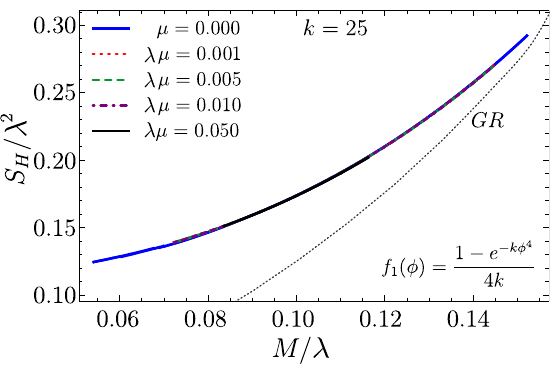}\hfill
          \includegraphics[scale=0.65]{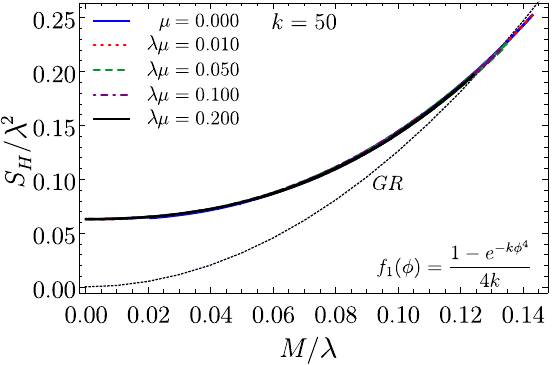}\\
          \includegraphics[scale=0.65]{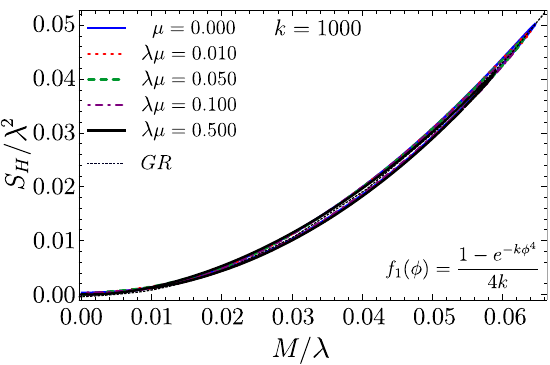}\\
		 \includegraphics[scale=0.65]{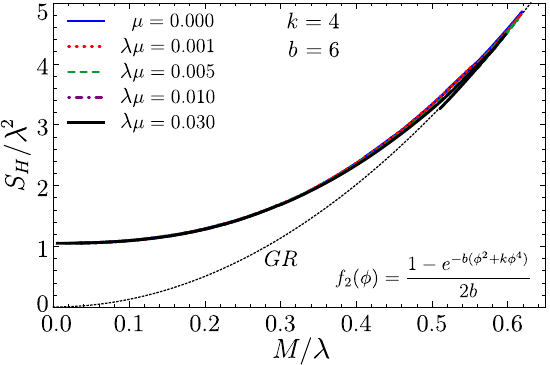}\hfill
   	 \includegraphics[scale=0.65]{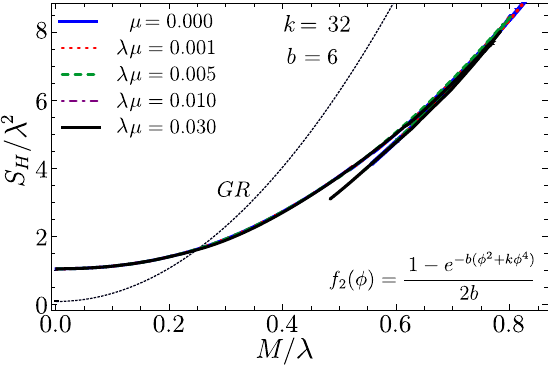}
	   \caption{Black hole's scaled entropy, $S_H/\lambda ^2$, as a function of the scaled ADM mass, $M/\lambda$, for several values of the scaled scalar field particle's mass, $\mu \lambda$, and five different cases of the scalarized GB black holes: nonlinear with coupling function \eqref{eq:coupling_nonline} and $k= 25$ (\textit{top-left}), $k=50$ (\textit{top-right}) and $k=1000$ (\textit{middle}); and mixed coupling function \eqref{eq:coupling_mixed} with $b=6$ and $k=4$ (\textit{bottom-left}) and $k=32$ (\textit{bottom-right}).}
	 	\label{F3}
		\end{figure}

%
    Consider now in more detail the scalarized black hole's thermodynamics. Both nonlinear and mixed scalarization endow black hole solutions that are thermodynamically preferable to the corresponding vacuum GR solution (see Fig.~\ref{F3} and \cite{doneva2018new,doneva2018charged}). Such is, however, not a generic feature.

\begin{figure}[h!]
		 \centering
		 \includegraphics[scale=0.65]{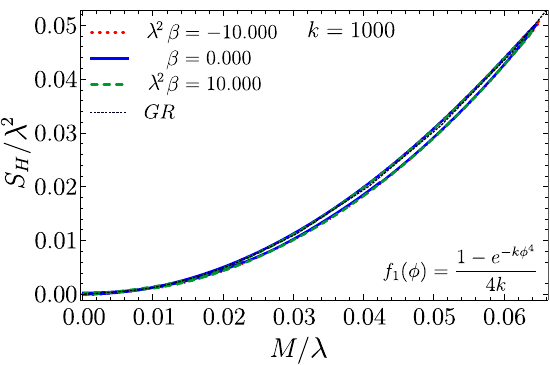}\hfill
          \includegraphics[scale=0.65]{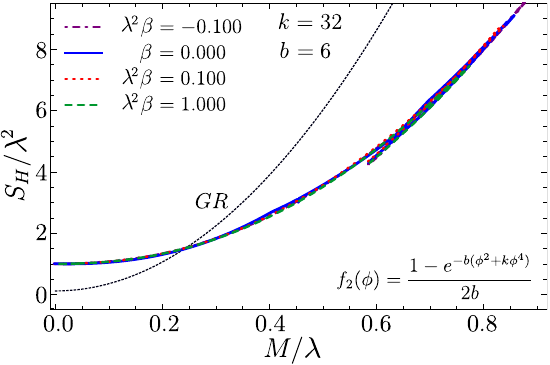}
	   \caption{Black holes scaled entropy, $S_H/\lambda ^2$,  as a function of the scaled ADM mass, $M/\lambda$, for several values of the scaled scalar field particle's quartic self-interaction, $\lambda^ 2 \beta$, and two different cases of the scalarized sGB black holes: a nonlinear with coupling function \eqref{eq:coupling_nonline} and $k=1000$ (\textit{left panel}); and a mixed coupling function \eqref{eq:coupling_mixed} with $b=6$ and $k=32$ (\textit{right panel}).}
	 	\label{F8}
		\end{figure}

	Let us start with the nonlinear scalarization (see Fig.~\ref{F3} top and middle, and Fig~\ref{F8} left). In the latter, for small enough values of $k\sim 25$, all the obtained solutions are entropically preferable when compared with vacuum GR. Increasing $k$ leads to a decrease in the entropy until an entropically unfavourable branch emerges. The addition of the mass/self-interaction term keeps the stable/unstable branch structure for high enough values of $k$ unchanged, reducing only the region of the parameter space for which each solution exists.

	In the case of mixed scalarization (see Fig.~\ref{F3} bottom), for small $k$ (bottom-left), the solutions contain a first entropically unfavourable branch (close to the bifurcation point from vacuum GR) and become entropically preferable at a second branch after the maximum of the mass. On the other hand, for high enough $k$ (bottom-right), only the very high value $\lambda$ solutions (small mass $M$) have an entropy higher than a comparable GR solution. The addition of the nonlinear coupling decreases the mixed scalarization entropy making them unfavourable.

    As in the case of nonlinear scalarization, the entropy structure for a given mixed scalarization $k$ is somewhat insensitive to the presence of the mass/self-interaction term. For a given $k$, the mixed scalarization entropy structure remains the same as one adds a mass/self-interaction term and no significant change in the point at which the solutions with high $k$ becomes thermodynamically unfavourable.

\begin{figure}[h!]
		 \centering
          \includegraphics[scale=0.65]{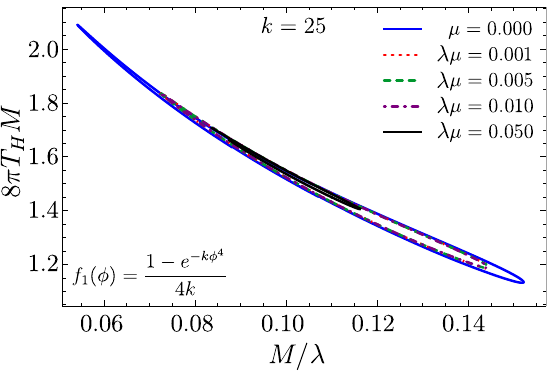}\hfill
          \includegraphics[scale=0.65]{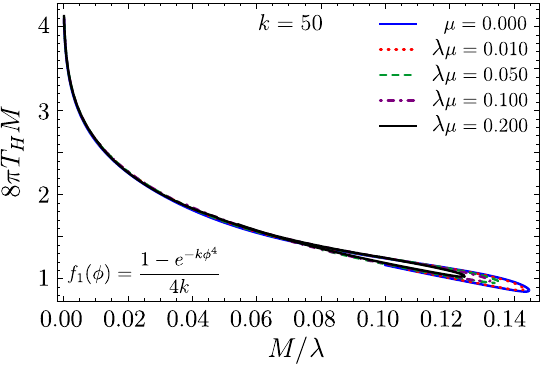}\\
          \includegraphics[scale=0.65]{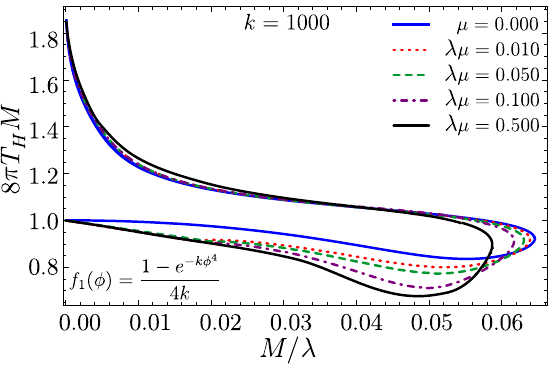}\\
		 \includegraphics[scale=0.65]{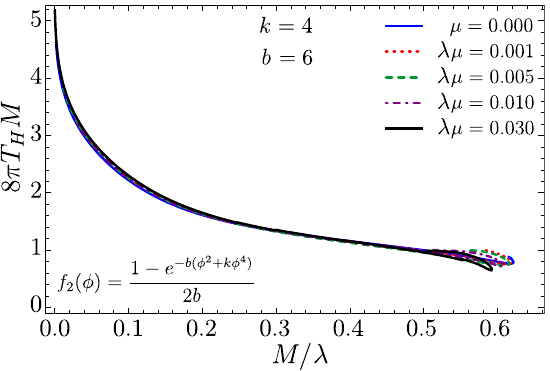}\hfill
          \includegraphics[scale=0.65]{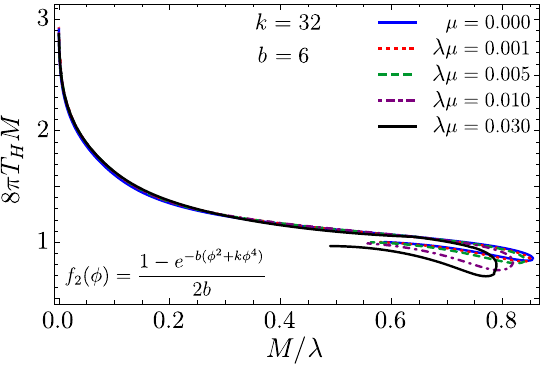}
	   \caption{Black holes normalized horizon temperature, $8\pi T_H M$, as a function of the scaled ADM mass, $M/\lambda$, for several values of the scaled scalar field particle's mass, $\mu \lambda$, and five different cases of the scalarized GB black holes: nonlinear with coupling function \eqref{eq:coupling_nonline} and $k= 25$ (\textit{top-left}), $k=50$ (\textit{top-right}) and $k=1000$ (\textit{middle}); and mixed coupling function \eqref{eq:coupling_mixed} with $b=6$ and $k=4$ (\textit{bottom-left}) and $k=32$ (\textit{bottom-right}).}
	 	\label{F4}
		\end{figure}

\begin{figure}[h!]
		 \centering
		 \includegraphics[scale=0.65]{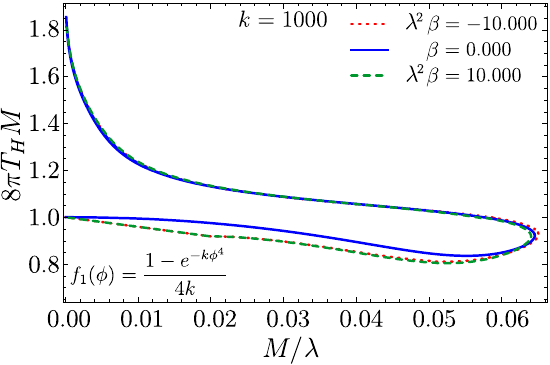}\hfill
          \includegraphics[scale=0.65]{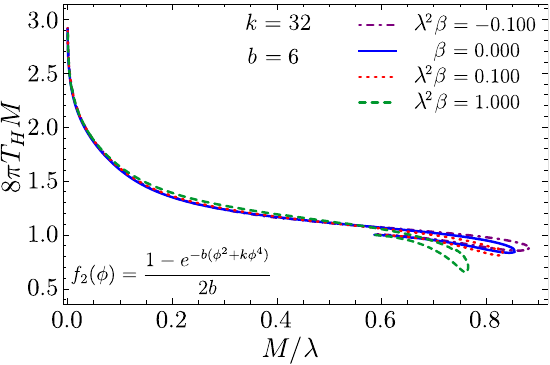}
	   \caption{Black holes normalized horizon temperature, $8\pi T_H M$, as a function of the scaled ADM mass, $M/\lambda$, for several values of the scaled scalar field particle's quartic self-interaction, $\lambda^ 2 \beta$, and two different cases of the scalarized sGB black holes: a nonlinear with coupling function \eqref{eq:coupling_nonline} and $k=1000$ (\textit{left panel}); and a mixed coupling function \eqref{eq:coupling_mixed} with $b=6$ and $k=32$ (\textit{right panel}).}
	 	\label{F9}
		\end{figure}

   Let us now study the normalized horizon temperature, $8\pi T_H M$, of both nonlinear and mixed scalarization in the presence of a massive, Fig.~\ref{F4}, or self-interacting, Fig.~\ref{F9}, scalar field.

    For small values of the nonlinear scalarization parameter $k$, the horizon temperature is, in general, higher than the Schwarzschild one (see Fig.~\ref{F4} top). The minimum value decreases as one increases the scalar field particle's mass (or positive self-interaction). 

    Nonlinearly scalarized black holes with high $k$ (middle) or mixed scalarization (bottom) have a region with lower temperature than the Schwarzschild black hole. The latter's width expands with the increase of the mass/positive self-interaction (the opposite occurs for the negative self-interaction, see Fig.~\ref{F9}). The minimum temperature in these regions also decreases with the increase of the mass/positive self-interaction without ever reaching zero.


  \bibliographystyle{ieeetr}
  \bibliography{main}

%
\end{document}